\documentclass[aps,prb,reprint,superscriptaddress,preprintnumbers,amsfonts,amssymb,amsmath,longbibliography]{revtex4-2}

\usepackage{graphicx}
\usepackage[dvipsnames]{xcolor}
\usepackage{hyperref}

\begin{document}

\title{Anomalously high quasiparticle thermal conductivity in the underdoped cuprate superconductor HgBa$_{2}$CuO$_{4+\delta}$}

\author{Jordan Baglo}
\email[]{Jordan.Baglo@USherbrooke.ca}

\author{Quentin Barth\'{e}lemy}

\author{\'{E}tienne Lefran\c{c}ois}

\affiliation{Institut Quantique, D\'{e}partement de physique \& RQMP, Universit\'{e} de Sherbrooke, Sherbrooke, Qu\'{e}bec, Canada}

\author{Anne Forget}
\author{Doroth\'{e}e Colson}
\affiliation{Service de Physique de l'\'{E}tat Condens\'{e}, CEA Saclay, France}

\author{Cyril Proust}
\affiliation{LNCMI-EMFL, CNRS UPR3228, Univ. Grenoble Alpes, Univ. Toulouse, INSA-T, Grenoble and Toulouse, France}
\affiliation{IRL Fronti\`{e}res Quantiques, Universit\'{e} de Sherbrooke--CNRS, Sherbrooke, Qu\'{e}bec, Canada}

\author{Louis Taillefer}
\email[]{Louis.Taillefer@USherbrooke.ca}
\affiliation{Institut Quantique, D\'{e}partement de physique \& RQMP, Universit\'{e} de Sherbrooke, Sherbrooke, Qu\'{e}bec, Canada}
\affiliation{IRL Fronti\`{e}res Quantiques, Universit\'{e} de Sherbrooke--CNRS, Sherbrooke, Qu\'{e}bec, Canada}
\affiliation{Canadian Institute for Advanced Research, Toronto, Ontario, Canada}

\date{\today}

\begin{abstract}
The single-layer cuprate superconductor HgBa$_{2}$CuO$_{4+\delta}$ (Hg1201) is an ideal candidate for investigating many properties of cuprates with minimal disorder and without the complication of multiple CuO$_2$ layers. Here we measure the in-plane longitudinal thermal conductivity $\kappa$ of underdoped Hg1201 ($T_c$ = 76 K, $p$ = 0.11) at dilution refrigerator temperatures to extract the nodal quasiparticle velocity ratio $v_F/v_\Delta$. Assuming contributions from only a single line node per quadrant on the Fermi surface leads to a value of $v_F/v_\Delta$ = $23 \pm 3$, anomalously large compared to other cuprates at similar dopings. In conjunction with the anomalously high quasiparticle specific heat of Hg1201 in the normal state reported previously at a similar doping, this points to more than one Fermi surface sheet crossing the nodal line, suggesting the presence of more than the single small electron pocket detected by quantum oscillations.
\end{abstract}
\maketitle

\section{Introduction}
After four decades of study, the high-$T_c$ cuprate superconductors continue to hold a prominent position at the forefront of condensed matter physics research. While much progress has been made, certain basic aspects of their physics remain difficult to pin down; in particular, the exact configuration of their Fermi surface, as well as the details of the competing order parameters believed to reconstruct that surface, remain hotly debated.

Among the techniques that can provide critical insight into these questions, thermal conductivity measurements in cuprates have played an important role. Unlike DC electrical transport and thermoelectric coefficients, thermal conductivity remains sensitive within the superconducting phase where electrical potentials are screened. As a bulk probe, it is also less sensitive to the surface quality issues that can hinder other techniques.

Perhaps the clearest case for the utility of thermal conductivity measurements in the cuprates came with the prediction of a universal limit for low-temperature conductivity of a nodal $d$-wave superconductor, as first predicted by Lee \cite{lee_localized_1993} and subsequently extended \cite{graf_electronic_1996,durst_impurity-induced_2000}. These works showed that, for sufficiently clean samples, the limiting low temperature $T$-linear contribution to thermal conductivity should be independent of quasiparticle lifetime due to scattering, and depend only on the gap anisotropy ratio $v_F/v_\Delta$ (where $v_F$ is the Fermi velocity and $v_\Delta \equiv \hbar^{-1}\left|\nabla_{\bf{k}}\Delta_{\bf{k}}\right|$ is the gap slope) and on the average spacing $n/c$ between superconducting planes \cite{durst_impurity-induced_2000}:
\begin{equation}
    \frac{\kappa_{0}}{T} \equiv \lim_{B,T \to 0}\frac{\kappa(B, T)}{T} = \frac{k_B^2}{3\hbar} \frac{n}{c}\left(\frac{v_F}{v_\Delta} + \frac{v_\Delta}{v_F}\right).
    \label{eq:univLimitKappaO}
\end{equation}
This expression assumes that a single Fermi surface crosses the nodal line only once per quadrant of the Brillouin zone, producing four nodal lines with Dirac-like quasiparticle dispersion; additional nodes coming from additional sheets in the Fermi surface would sum accordingly. The utility here is the ability to extract a quantitative value for $v_F/v_\Delta$ for comparison with other techniques, from $\kappa_{0}/T$ and unit cell dimensions alone.

Such low-temperature measurements were reported soon after, starting with YBa$_2$Cu$_3$O$_{6+x}$ (YBCO) \cite{taillefer_universal_1997,chiao_quasiparticle_1999} and Bi$_2$Sr$_2$CaCu$_2$O$_8$ (Bi2212) \cite{chiao_low-energy_2000,nakamae_effect_2001}, confirming the insensitivity of the experimentally observed residual linear term $\kappa_{0}/T$ to moderate levels of disorder and a gap ratio consistent with independent measurements from other techniques, including Angle-Resolved Photoemission Spectroscopy (ARPES). The confirmation of this universal behavior reinforced confidence in the applicability of the theoretical framework of $d$-wave superconductivity to the high-$T_c$ cuprates. Subsequent thermal conductivity work in YBCO \cite{sutherland_thermal_2003, hill_transport_2004} and Tl$_2$Ba$_2$CuO$_{6+\delta}$ (Tl2201) \cite{hawthorn_doping_2007} extended this work more broadly across the phase diagram. Note that for most other cuprates (including {La$_{2-x}$}{Sr$_x$}{Cu}{O$_{4+\delta}$}), sufficiently clean single crystals to access the universal conductivity regime have not yet been available \cite{hawthorn_field-induced_2003}.

One prominent omission from such studies to date has been the single-layer cuprate HgBa$_{2}$CuO$_{4+\delta}$ (Hg1201), which is notable for possessing the highest-known $T_c$ at optimal doping (97 K) of all single-layer cuprates, as well as being one of the cleanest cuprates available \cite{barisic_demonstrating_2008}. With its simple tetragonal structure and the absence of many of the particularities which complicate the study of other cuprates---such as multiple layers, CuO chains, structural distortions, and buckling---it is sometimes described as a ``model'' cuprate \cite{barisic_demonstrating_2008}. Hg1201 is also relevant as the $n = 1$ end member of the HgBa$_2$Ca$_{n-1}$Cu$_n$O$_{2(n+1)+\delta}$ family of $n$-layer Hg-based cuprates (Hg1212 and Hg1223 for $n =$~2 and 3, respectively). Hg1223 is the superconductor with the highest $T_c$ of all known materials (at ambient pressure). In this respect, we strive to first understand the single-layer Hg1201 as a foundation for approaching its multi-layer relatives.

The Fermi surface of Hg1201 and the nature of its reconstruction have been explored with several techniques, all for dopings in the vicinity of $p=0.1$. Quantum oscillation measurements \cite{barisic_universal_2013, chan_single_2016} revealed the presence of a single frequency, ascribed to an electron pocket. Hall  \cite{doiron-leyraud_hall_2013, chan_extent_2020} and Seebeck \cite{doiron-leyraud_hall_2013} measurements display a change in sign from positive to negative at temperatures below $\sim{}20$~K, indicative of a Fermi surface reconstruction from hole-like to electron-like at low temperature. Evidence for short-range incommensurate charge density wave correlations at $p \approx 0.1$ has been found both in X-ray scattering \cite{tabis_charge_2014, tabis_synchrotron_2017} and NMR experiments \cite{lee_coherent_2017}. For most authors, the presence of an electron pocket in the Fermi surface is attributed to charge order \cite{proust_taillefer_2019}. However, the effects of this reconstruction on the nodal dispersion, and the possibility of additional pockets in the Fermi surface (besides the small electron pocket) have remained unexplored.

Specific heat measurements by Girod \textit{et al.} on Hg1201 at $p = 0.09$ \cite{girod_high_2020} revealed a residual $T$-linear specific heat coefficient in the normal state (at high field) that is three times greater than that expected from a single electron pocket (with a mass as measured by quantum oscillations). This showed that additional Fermi surface sheets must be present, making underdoped Hg1201 a promising material for low-temperature thermal conductivity studies, since these would be sensitive to the presence of any such sheets, were they to cross the nodal line.

\section{Methods}

The Hg1201 sample was synthesized using a self-flux growth technique as described in \cite{legros_crystal_2019}. The critical temperature $T_c = 76\pm{}1$~K was measured by vibrating sample magnetometry at 100~Oe in a Quantum Design PPMS, corresponding to a hole doping of $p=0.109 \pm 0.005$ using the $T_c(p)$ relationship determined in \cite{yamamoto_thermoelectric_2000}. Gold contact pads were sputtered onto the sample followed by a high-temperature anneal, to which 25-\textmu{}m silver wire leads were attached using DuPont 4929N silver paint. The sample region where the thermal gradient was measured was of roughly rectangular cross-section, 95 \textmu{}m thick and 710 \textmu{}m wide, with a length of 470 \textmu{}m between the longitudinal temperature contacts. Thermal transport measurements for this sample at higher temperatures were previously presented in \cite{altangerel_thermal_2025}.

Longitudinal thermal conductivity measurements were performed in a $^3$He/$^4$He dilution refrigerator using a conventional one-heater--two-thermometer configuration \cite{sutherland_thermal_2003, hawthorn_doping_2007}, with heat current applied along an $a$ axis in plane to generate a thermal gradient of $dT/T \approx 3\%$. The sample was anchored at one end with silver paint to a copper block heat sunk to the probe, with the other end connected via silver wire to a resistive heater as the thermal current source. Sample temperatures were measured using ruthenium oxide chip resistors thermally connected to the sample via Ag wires 25 \textmu{}m in diameter; both thermometers and heaters are otherwise thermally isolated by suspension from nylon posts with Kevlar threads, using resistive Pt$_{0.92}$W$_{0.08}$ coils for electrical connections. 

Measurements were taken for several static values of magnetic field applied out of plane ($B \parallel \hat{c}$) from 0 T to 15 T, in small sample temperature steps from $\sim{}\!80$~mK to at least 1 K. Consecutive heat-on/heat-off measurements were taken at each temperature to allow for \textit{in situ} calibration of each thermometer in field with reference to a calibrated Ge thermometer in a field-compensated zone. All magnetic field changes were performed with the sample temperature held above 120 K, well in excess of its 76 K critical temperature.

\section{Results and Discussion}
\begin{figure}
\includegraphics[width=\columnwidth]{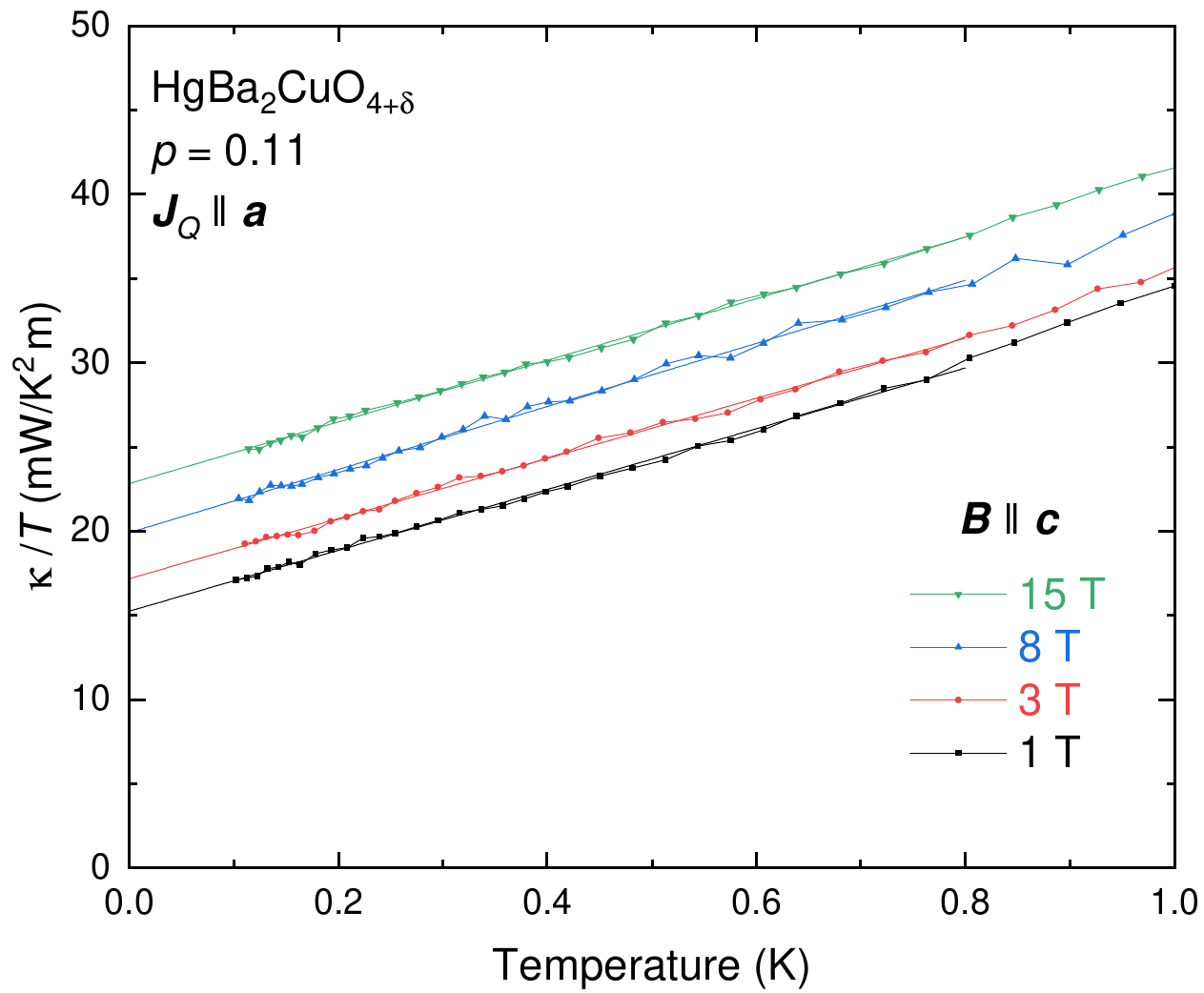}
\caption{\label{fig:linKappaOverT}Low temperature thermal conductivity of Hg1201 at $p=0.11$ measured at several magnetic field values as indicated, plotted as $\kappa/T$ vs $T$. The observed linear low-temperature behavior is used to extrapolate to $T = 0$ (solid lines) to provide the $\kappa_0(B)/T$ data shown in Fig.~\ref{fig:KHfits}.}
\end{figure}

\begin{figure}
\includegraphics[width=\columnwidth]{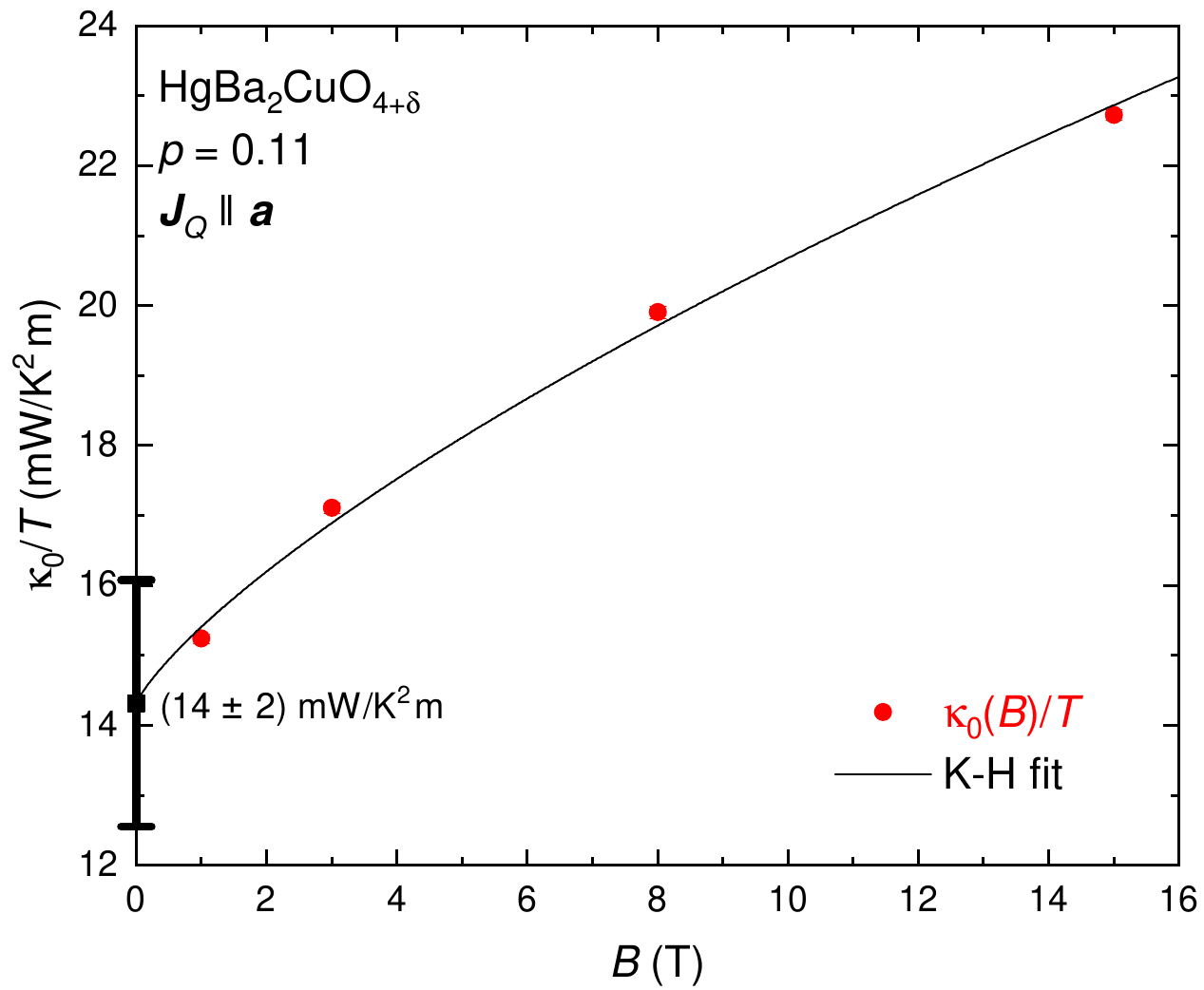}
\caption{\label{fig:KHfits}Extrapolated $T\rightarrow{}0$ thermal conductivity data $\kappa_0(B)/T$ resulting from the fits in Fig.~\ref{fig:linKappaOverT}, plotted as $\kappa_0/T$ vs $B$. The solid line is a fit to the K\"{u}bert-Hirschfeld model~\cite{kubert_quasiparticle_1998} (Eq.~\ref{eq:KubHirsch1998}). The resulting extrapolation to $B=0$ gives $\kappa_{0}/T = 14 \pm 2$~mW/K$^2$m (this error includes the uncertainty in sample geometry).}
\end{figure}

The resulting measurements are presented as $\kappa/T$ vs.\ $T$ in Fig.~\ref{fig:linKappaOverT}. The $\kappa(B,T)/T$ data are fit linearly in $T$ from 0.1 to 0.8 K to infer the limiting low-temperature electronic contribution $\kappa_0(B)/T$, shown vs field as the points in Fig.~\ref{fig:KHfits} \footnote[1]{Note that a fit to a free power law $A + B T^y$ gives a shared exponent of $y = 0.96 \pm 0.02$, very nearly linear; fitting these alternative extrapolated $\kappa_0(B)/T$ to Equation \ref{eq:KubHirsch1998} gives identical values (within quoted precision) of $\kappa_{0}/T \equiv A$, and therefore $v_F/v_\Delta$, to the linear case ($y = 1$).}.
While $\kappa(B,T)/T$ curves at fixed finite magnetic fields $B$ can clearly be extrapolated linearly to $T = 0$, a low-temperature downturn for $B = 0$ (likely an artifact of electron-phonon decoupling \cite{smith_origin_2005}) renders the resulting extrapolation somewhat ambiguous (see Appendix for more details). Instead, we extrapolate our finite-field data to $B=0$ (Fig.~\ref{fig:KHfits}), accounting for the field dependence using the model of K\"{u}bert and Hirschfeld \cite{kubert_quasiparticle_1998}, which attributes the $B$ dependence of $\kappa_0/T$ to the Volovik effect \cite{volovik_effect_1993} (quasiparticle Doppler shift due to the superfluid flow in a vortex lattice):
\begin{equation}
        \frac{\kappa_0(B)}{\kappa_{0}(B\rightarrow 0)} = \frac{\rho^2}{\rho\sqrt{1+\rho^2}-\sinh^{-1}\rho},
        \label{eq:KubHirsch1998}
\end{equation}
where $\rho \equiv \frac{\gamma_B}{a\hbar v_F}\sqrt{\frac{3\Phi_0}{B}}$; here $\gamma_B$ is the impurity bandwidth, $a$ is a vortex lattice geometry factor of order unity, $v_F$ is the Fermi velocity, and $\Phi_0 \equiv h/2e$ is the flux quantum. Here we take $a=1/2$ determined for a square lattice \cite{hussey_low-energy_2002,Note2}.
\footnotetext[2]{From Ref.~\cite{hussey_low-energy_2002}, for a square lattice $a=0.5$ and for a triangular lattice $a=0.465$; NMR measurements of underdoped Hg1201 \cite{lee_magnetic-field-induced_2017} found an oblique vortex lattice of angle $\alpha = 73 \pm 7^\circ{}$ in the relevant field range.}

From the resulting fit shown in Fig.~\ref{fig:KHfits}, we extract a value for the universal conductivity of $\kappa_{0}/T = 14 \pm 2$~mW/K$^2$m. Assuming that the single CuO$_2$ layer contributes only one nodal line per Brillouin zone quadrant ($n = 1$), Eq.~\ref{eq:univLimitKappaO} yields a value $v_F/v_\Delta = 23~\pm~3$.

The applicability of Eq.~\ref{eq:univLimitKappaO} only holds up to a certain level of scattering. Several pieces of experimental evidence point to a sufficiently high degree of cleanliness: Thermal Hall conductivity measurements on the exact same Hg1201 sample \cite{altangerel_thermal_2025}, and the observation of quantum oscillations in similar Hg1201 samples \cite{barisic_universal_2013}, including some grown by the same group \cite{oliviero_magnetotransport_2022}. Measurements of the Nernst coefficient, which is proportional to carrier mobility, found similar magnitudes for Hg1201 and YBCO \cite{doiron-leyraud_hall_2013}, with the latter being renowned for its cleanliness.

From the fits to Eq.~\ref{eq:KubHirsch1998} shown in Fig.~\ref{fig:KHfits}, we infer an impurity bandwidth $\gamma_B$~=~5.3~meV (given $a = 1/2$ and using $v_F=2.5\times{}10^5$~m/s \cite{altangerel_thermal_2025}). Assuming unitary limit scattering with a normal state scattering rate $\Gamma$, for which $\gamma_B \simeq 0.61 \sqrt{\hbar \Gamma \Delta_0}$ \cite{kubert_quasiparticle_1998}, this yields $\frac{\hbar\Gamma}{\Delta_0} \approx 0.12$, assuming $\Delta_0=4 k_{B}T_c = 25$~meV. Previous thermal conductivity measurements of Zn-doped YBCO \cite{taillefer_universal_1997} as well as Sr$_2$RuO$_4$ \cite{suzuki_universal_2002, hassinger_vertical_2017}---both being superconductors with vertical line nodes---have demonstrated only weak dependence of the residual linear term $\kappa_{0}/T$ on the scattering rate $\Gamma$, at least up to a value of $\hbar \Gamma/k_B T_c$ comparable to that found here. We conclude that the level of scattering in our Hg1201 sample is sufficiently small for Eq.~\ref{eq:univLimitKappaO} to provide a valid determination of $v_F/v_\Delta$.

\begin{figure}
\includegraphics[width=\columnwidth]{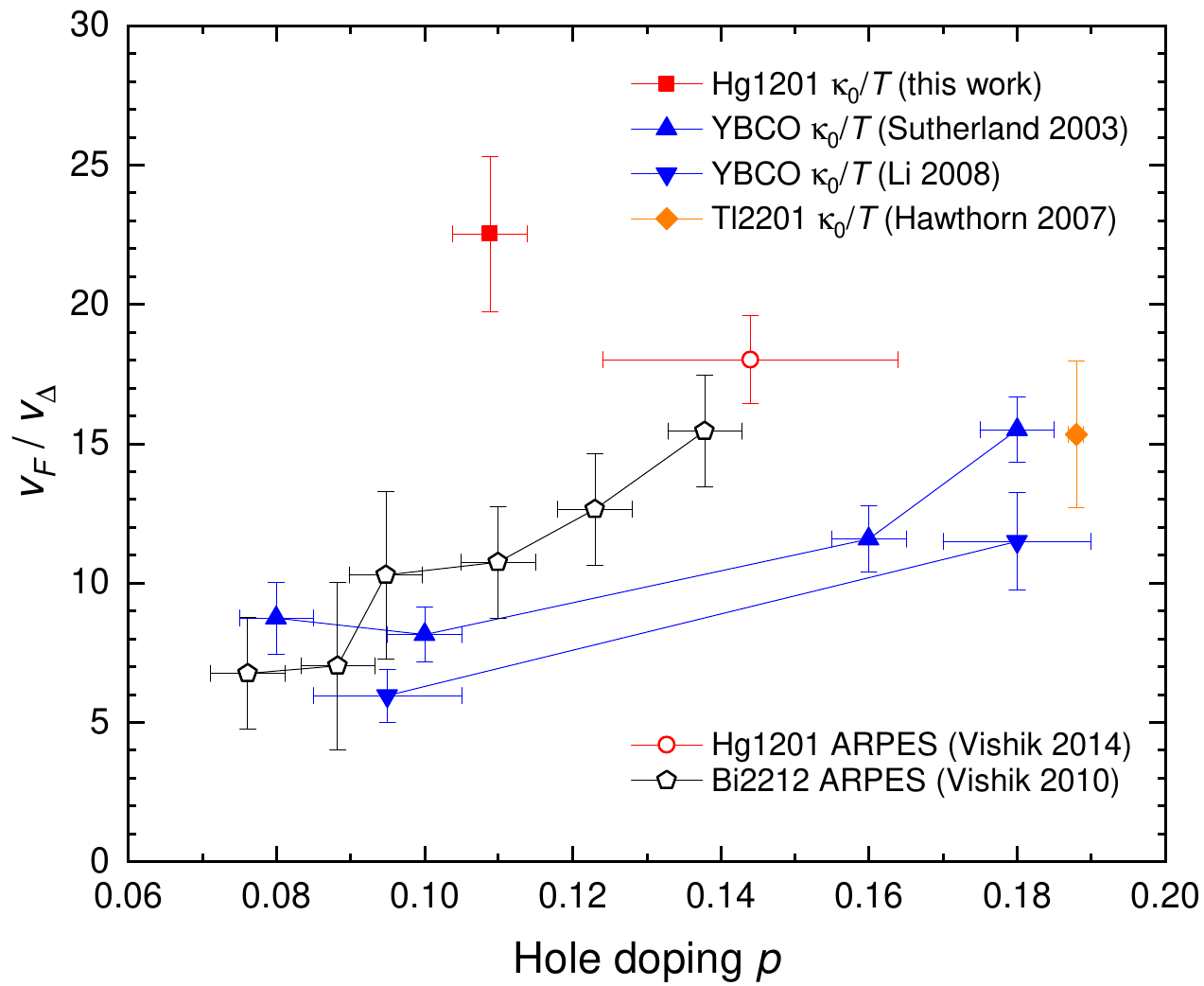}
\caption{\label{fig:vFOvervDcompare}Comparison of $v_F/v_\Delta$ for Hg1201 at $p = 0.11$ (red square, this work) with corresponding values obtained for two other cuprates: YBCO (blue triangles \cite{sutherland_thermal_2003, li_low-temperature_2008}) and Tl2201 (orange diamond \cite{hawthorn_doping_2007}). This is compared to estimates from ARPES for Hg1201 (red circle \cite{vishik_angle-resolved_2014}) and Bi2212 (black pentagons \cite{vishik_doping-dependent_2010}). We see that $v_F/v_\Delta$ inferred for Hg1201---assuming line nodes cross $n = 1$ Fermi sheets per original Brillouin zone quadrant---is found to be significantly higher than for other cuprates at similar dopings (where $n=1$ is also assumed).}
\end{figure}

In Fig.~\ref{fig:vFOvervDcompare}, we compare our value of $v_F/v_\Delta$ with values measured for other cuprates as a function of doping. We can readily see that the value we obtain here for Hg1201 is significantly larger than that of similar compounds near $p=0.11$ for which reliable estimates are available. Included here (in legend order) are thermal conductivity results for YBCO from two different studies \cite{sutherland_thermal_2003,li_low-temperature_2008} and for Tl2201 \cite{hawthorn_doping_2007}, as well as ARPES results for Hg1201 \cite{vishik_angle-resolved_2014} and Bi2212 \cite{vishik_doping-dependent_2010}.

Such an unusually large ratio $v_F/v_\Delta$ compared to other hole-doped cuprates is surprising. Note that from the larger effective masses seen in quantum oscillation measurements of Hg1201 ($m^* = 2.7\,m_0$) compared to YBCO ($m^* = 1.9\,m_0$) one might instead expect a \emph{smaller} value of the Fermi velocity (and thus a smaller $v_F/v_\Delta$) for Hg1201. Some prior theoretical work has proposed scenarios where universal conductivity may be violated \cite{atkinson_optical_2002,gusynin_thermal_2004,andersen_breakdown_2008,schiff_effect_2010}, but their applicability to the case of Hg1201 is unclear.

Ultimately, we are left with a large residual $T$-linear thermal conductivity $\kappa_{0}/T$ that is difficult to explain starting from a scenario in which the Fermi surface contains only a single electron-like pocket hosting gap nodes. There exists strong experimental support for additional Fermi surface sheets in Hg1201: our large value of $\kappa_0/T$ in Hg1201 at $p = 0.11$ is highly reminiscent of the specific heat measurements from Girod \textit{et al.} on Hg1201 at $p = 0.09$ \cite{girod_high_2020}, where they observed a residual $T$-linear specific heat coefficient $\gamma = 12 \pm 2$ mJ/K$^2\,$mol in the normal state, markedly greater than the $\sim{}5$~mJ/K$^2\,$mol found for similar cuprates near that doping (\cite{girod_high_2020}, and references therein). Moreover, the $\gamma$ value expected from a Fermi surface that consists only of a single electron pocket (with $m^* = 2.7\,m_0$ \cite{chan_single_2016}) is $\gamma = 3.7 \pm 0.2$ mJ/K$^2\,$mol \cite{girod_high_2020}, 3 times lower than the measured value (and much closer to other cuprates). To account for an observed normal-state $\gamma$ more than 3 times greater, the authors suggested the presence of additional (as-yet-unobserved) Fermi sheets whose contributions would make up the balance.

This would also provide a natural resolution here: additional Fermi surface sheets with line nodes would give additional nodal quasiparticle contributions to $\kappa_{0}/T$. Starting from a single node ($n = 1$) with a value of $v_F/v_\Delta = 10 \pm 2$ typical of cuprates at the same doping (Fig.~\ref{fig:vFOvervDcompare}), one would expect a contribution to $\kappa_{0}/T$ of 6.4~$\pm$~1.3~mW/K$^2$m. The remaining 7.9~$\pm$~1.8~mW/K$^2$m would correspond to that expected from a single additional nodal Fermi surface with $n = 1$ and $v_F/v_\Delta = 12 \pm 3$. (Alternatively, two identical nodal pockets with $v_F/v_\Delta = 11.3 \pm 1.4$ would also account for the measured $\kappa_{0}/T$.)

This possibility is not without motivation; several theoretical scenarios have been put forth for generating additional Fermi surface pockets and additional nodal points, such invoking Fermi surface reconstruction by charge order, and others proposing more exotic means such as fractionalization. Of the former category, the most prominent Fermi surface reconstruction scenario at present---that of a biaxial charge density wave \cite{harrison_protected_2011} of two wavevectors $(Q_\mathrm{CDW},0)$ and $(0,Q_\mathrm{CDW})$, with $Q_\mathrm{CDW}\approx 0.275$---should produce a folded Brillouin zone including two additional small hole-like pockets \cite{tabis_charge_2014,chan_single_2016,tabis_synchrotron_2017}. While possible evidence for such hole pockets has sometimes been claimed for YBCO \cite{allais_connecting_2014,doiron-leyraud_evidence_2015}, for Hg1201 these pockets have not yet been observed by quantum oscillation measurements nor by ARPES. Such additional pockets would also not cross the nodal line and thus would not be expected to generate the additional nodal contributions needed to explain an enhanced residual $\kappa_{0}/T$.

An additional issue with this interpretation stems from the short-range nature of the 2D charge density fluctuations seen at low field in Hg1201; it has been pointed out \cite{uchida_ubiquitous_2021} that the correlation lengths $\xi$ of such fluctuations are in fact the shortest observed among the cuprates, with values of $\xi \approx 20$\AA{} only spanning $\sim 5$ unit cells, only marginally longer than the $\sim 3.5$ unit cell modulation period \cite{tabis_charge_2014, tabis_synchrotron_2017}. It is not clear that such short-range charge density fluctuations would be sufficient to effectively reconstruct a Fermi surface and generate new pockets \cite{gannot_fermi_2019}. Any appropriate resolution must also explain the difference between Hg1201 and YBCO: na\"i{}vely, the effects of charge order would be expected to only be stronger for YBCO, for which no similarly enhanced $\kappa_{0}/T$ \cite{sutherland_thermal_2003,li_low-temperature_2008} (or normal-state $\gamma$ \cite{michon_thermodynamic_2019, kacmarcik_unusual_2018}) has been measured.

Other similar ``reconstruction-type'' scenarios have been proposed, including spin-density wave reconstruction of a Bogoliubov Fermi surface resulting from loop-current order\cite{allais_loop_2012}, and the Fermi surface of a fractionalized Fermi liquid reconstructed by charge order \cite{senthil_fractionalized_2003, bonetti_quantum_2024}, both of which \emph{do} result in finite electron spectral weight crossing a nodal line and contributing to the low-energy density of states (and thus $\kappa_{0}/T$). However, both scenarios would be subject to the same questions regarding the short correlation lengths observed for charge (or spin) order in Hg1201.

More recent work on the fractionalized Fermi liquid framework as applied to the cuprate pseudogap \cite{bonetti_fractionalized_2026} emphasizes the existence of closed hole pockets with completed backsides, with no need to invoke reconstruction by Brillouin zone folding due to order. Recent $c$-axis high-field angle-dependent magnetoresistance measurements of $T_c$=74~K Hg1201 at 85 K \cite{chan_observation_2025} have shown evidence (through the observation of Yamaji peaks \cite{yamaji_angle_1989}) of small Fermi pockets each occupying roughly 1.3\% of the Brillouin zone area, compared to $\sim{}\!3\%$ each from quantum oscillation studies at low temperatures \cite{barisic_universal_2013, chan_single_2016, chan_extent_2020, oliviero_magnetotransport_2022, oliviero_charge_2024}; this has been argued \cite{zhao_yamaji_2026} to be consistent with the Fermi surface predicted for the fractionalized Fermi surface model \cite{bonetti_fractionalized_2026} which would yield hole-like pockets with areas corresponding roughly to $p/8$. While this model would be expected to produce additional nodal contributions from the closed backsides of the hole pockets, it is claimed that these additional nodes may annihilate with spinons, leaving only the original four line nodes. Additionally, given the evidence from transport measurements \cite{doiron-leyraud_hall_2013, chan_extent_2020} that the Fermi surface of Hg1201 would be reconstructed at a temperature well below that of the $T=85$~K Yamaji angle work, it is unclear to what extent the electronic structure measured at $T = 85$~K differs from the low-temperature regime relevant in this work.

\section{Conclusion}
Our thermal conductivity measurements on Hg1201 at $p = 0.11$ have revealed a large residual thermal conductivity $\kappa_{0}/T$ that is difficult to reconcile with a Fermi surface consisting of a single electron pocket. While the details of a compatible reconstruction scenario need to be clarified, both our data and prior specific heat data (in the normal state) are most naturally explained by invoking additional sheets in the Fermi surface of Hg1201 hosting gap nodes.

Because Hg1201 is a cuprate superconductor in its simplest form (no chains, nor multiple layers, nor structural distortion) and with a high level of order, our findings in principle reflect a basic characteristic of cuprates that should apply generally, \emph{modulo} any complication due to deviations from the ideal cuprate structure. Having said this, the detailed Fermi surface of various hole-doped cuprates may well differ from one material to another if the charge modulations that cause its reconstruction differ, whether in their period, amplitude, or correlation length.

\appendix*
\section{Temperature dependence of thermal conductivity at low magnetic field}
\begin{figure}
\includegraphics[width=\columnwidth]{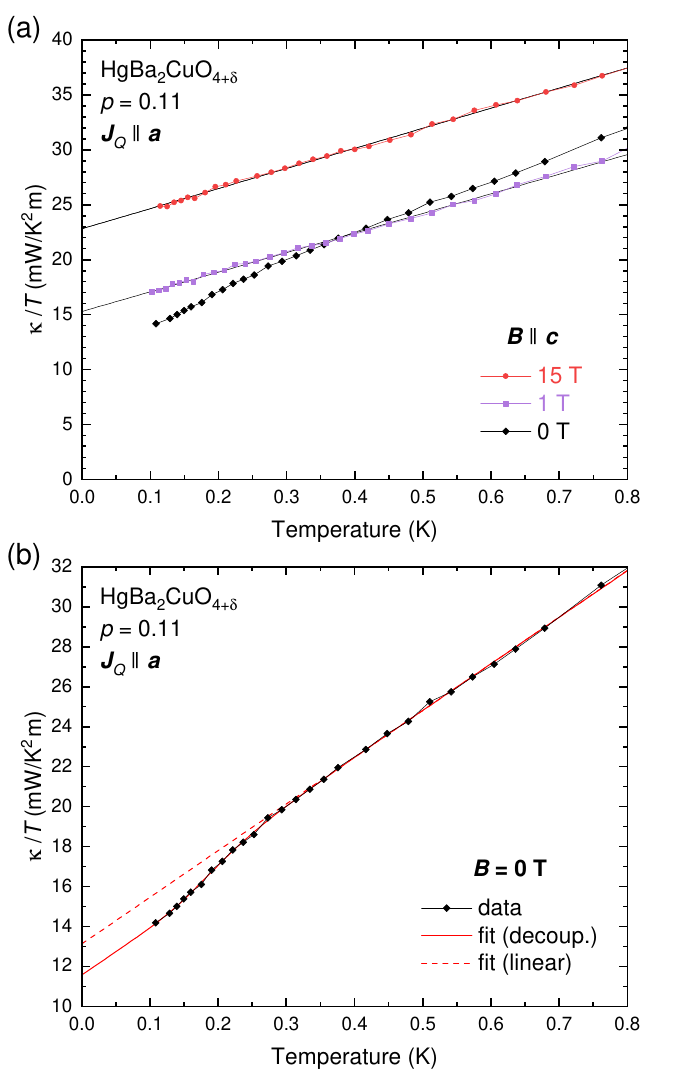}
\caption{\label{fig:lowTkappa}(a) Low temperature thermal conductivity of Hg1201 at $p=0.11$ measured at three magnetic field values, including $B = 0$~T (black points), plotted as $\kappa/T$ vs $T$; lines are linear fits in the interval from 0.1 to 0.8~K. (b) Zoom on the zero-field data: The clear low-temperature downturn suggests the onset of electron-phonon decoupling, here fit to Eq.~\ref{eq:decoup} (red solid line), with a simple linear fit shown for comparison (red dashed line); for higher fields this decoupling is not discernible down to 100 mK (see panel (a) and Fig.~\ref{fig:linKappaOverT}).}
\end{figure}

Thermal conductivity data were collected at several values of magnetic field, including $B=0$~T. As illustrated in Fig.~\ref{fig:lowTkappa}(a), there was a qualitative difference between the zero-field data (black points)---showing a downturn below 0.3~K---and those taken in finite applied field---which remain linear down to 0.1~K.

As explained in Ref.~\cite{smith_origin_2005}, at low temperatures the thermal transfer between electrons (or quasiparticles) and phonons weakens dramatically, so that when contacts to the sample have poor electrical conductance, electrons will no longer be in thermal equilibrium with the thermometers. For typical sample geometries and parameters, this manifests as a rapid downturn in \emph{apparent} thermal conductivity below a decoupling temperature $T_\mathrm{dec}$. As derived in Ref.~$\cite{smith_origin_2005}$, we start with a ``true'' thermal conductivity of the form $\kappa/T = \alpha + \beta T$, with $\kappa_{el}/T = \alpha$ representing the low-temperature electronic residual term and $\kappa_{ph}/T = \beta T$ the phononic contribution (with predominant scattering from electrons---see \cite{hawthorn_doping_2007}). In the presence of thermal decoupling, the measured conductivity may be modeled by
\begin{equation}
 \frac{\kappa}{T} = \alpha \frac{1}{1+\frac{r}{1+r(T/T_{\mathrm{dec}})^{n-1}}}+\beta T,
 \label{eq:decoup}
\end{equation}
where $r$, $T_\mathrm{dec}$, and $n$ depend on sample and contact details. In this model the apparent $\kappa_{el}/T$ drops from $\alpha$ above $T_\mathrm{dec}$ to $\alpha/(1+r)$ as $T\rightarrow 0$. The red solid curve in Fig.~\ref{fig:lowTkappa}(b) is a fit to Eq.~\ref{eq:decoup} of the $B=0$~T data up to 0.8~K; the curvature at low temperature is described well, with $T_\mathrm{dec} = 0.15$~K. For finite fields no such downturn is evident down to 100 mK (Figs.~\ref{fig:linKappaOverT} and \ref{fig:lowTkappa}), consistent with lower decoupling temperatures $T_\mathrm{dec} \lesssim 60$~mK. Direct evidence that electrons and phonons are more weakly coupled at $B=0$ is the fact that the $\kappa/T$ curve at $B=0$ exceeds the $B=1$~T curve at temperatures above $\sim{}\!0.4$~K (Fig.~\ref{fig:lowTkappa}(a)).

Note that a linear fit to the $\kappa/T$ vs $T$ curve at $B=0$ above 0.25~K yields $\kappa_0(0)/T = 13$~mW/K$^2$m (see dashed line in Fig.~\ref{fig:lowTkappa}(a)). This is consistent, within error bars, with the value obtained by extrapolating the in-field values of $\kappa_0(B)/T$ to $B\rightarrow0$, as shown in Fig.~\ref{fig:KHfits}.

\begin{acknowledgments}
We thank S.~Fortier for his assistance with the experiments, and the cryogenics team at the Institut Quantique
for their support.
L.\,T. acknowledges support from the Canadian Institute for Advanced Research (CIFAR) as a CIFAR Fellow, 
and funding from the Institut Quantique,
the Natural Sciences and Engineering Research Council of Canada (Grant No.~PIN:123817),
and a Canada Research Chair.
C.\,P. acknowledges support
from the EUR grant NanoX No.~ANR-17-EURE-0009 and
from the ANR grant FRONTQUANT No.~ANR-24-CE97-0004.
J.\,B., Q.\,B., {\'E}.\,L., and L.\,T. have benefited from their affiliation to the RQMP~\cite{RQMP}.
This research was undertaken thanks, in part, to funding from the Canada First Research Excellence Fund.
This research project No.~324046 is made possible thanks to funding from the Fonds
de recherche du Qu\'{e}bec.

\end{acknowledgments}


%

\end{document}